\input{aipcheck}

\documentclass[
    ,final            
  ]
  {aipproc}

\layoutstyle{8x11double}

\begin{document}

\title{The Sun as a Star: 13 years of SoHO}


\author{R.A. Garc\'\i a}{
  address={Laboratoire AIM, CEA/DSM-CNRS-Universit\'e Paris Diderot; CEA, IRFU, SAp, centre de Saclay, F-91191, Gif-sur-Yvette, France}
}

\begin{abstract}
The best known Solar oscillation-like star is the Sun. During the last 14 years, the ESA/NASA Solar and Heliospheric Observatory (SoHO) has been continuously observing this star from the Lagrange point L1 with an enormous success. Among the 11 instruments placed onboard, 3 of them are dedicated to helioseismology: GOLF, VIRGO and MDI. The first two observe the Sun as a star by integrating the velocity or intensity signal of the visible solar disk into a single point. They are thus similar to any other observation done in asteroseismology. During this review I will present the most important results obtained during the mission concerning the Sun seen as a star and how this results have evolved our thoughts of the inside of our star.
 \end{abstract}

\maketitle


\section{Introduction}
The Solar and Heliospheric Observatory (SoHO) is a three-axis, stabilized spacecraft developed by the European Space Agency (ESA) in collaboration with the National Aeronautics and Space Administration (NASA). It contains eleven scientific instruments dedicated to the study of the Sun, its heliosphere and the solar wind \citep{DomFle1995}. SoHO was one of the cornerstones of the ESA Space Science program called Horizon 2000 and it was succesfully launched from the Kennedy space flight center on December 2, 1995 at 8:08 GMT by an Atlas Centaur II As rocket.  

SoHO offers an unprecedented opportunity to study the deep interior of the Sun through helioseismology under ideal conditions at the Lagrange $L_1$ point at 1.5 10$^6$ km from Earth. At this location, no terrestrial atmospheric effects are present, continuous exposures to the Sun are possible (more than 95 $\%$ duty cycle), and there is a low Sun-spacecraft line-of-sight velocity. This spacecraft carries three Helioseismic instruments: GOLF\footnote{Global Oscillations at Low Frequency\citep{GabGre1995}}, SOI/MDI\footnote{Solar Oscillation Imager/Michelson Doppler Interferometer\citep{1995SoPh..162..129S}}, and VIRGO\footnote{Variability of solar Irradiance and Gravity Oscillations\citep{1995SoPh..162..101F}}. 
  
In this review, I will concentrate on Sun-as-a-star observations because they are similar to what we can do in asteroseismology today, reducing the modes that can be detected to those with degree l $\le$ 4. Low-degree modes have been very important to better constrain the solar models and improve our knowledge of the solar interior (see e.g. \citep{STCBas1997,STCCou2001,2004PhRvL..93u1102T,2008SoPh..251...53C}). I will focus here on a few results that can be interesting for asteroseismic investigations and that have been obtained by GOLF and VIRGO but without forgetting the ground-based BiSON\footnote{Birmingham Solar-Oscillations Network \citep{1996SoPh..168....1C}} network which started working -- with a full coverage of the Sun -- at the same epoch when SoHO started operations. Of course, we cannot forget that imaged instruments --as MDI and the ground based network GONG\footnote{Global Oscilation Network Group \citep{HarHil1996}}-- are required to compute more accurate inversions and thus, better infer quantities in the radiative region like the rotation \citep{CouGar2003,GarCor2004} or the sound speed \citep{2009ApJ...699.1403B}. 

\section{Sun-as-a star observations: The PSD}
\begin{figure}[!htb]
 \includegraphics[height=.3\textheight, width=0.481\textwidth]{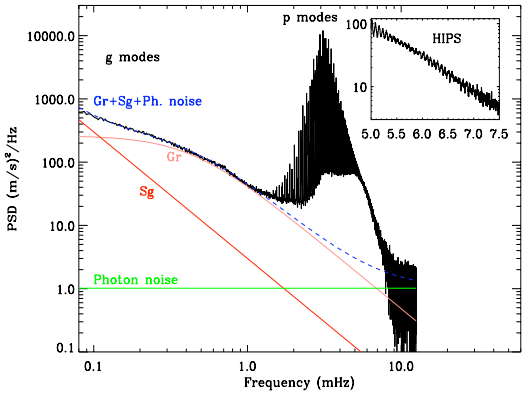}
  \caption{\label{fig1} Power spectrum density of the GOLF instrument. We have also plotted a Harvey-type fitting of the convective background with two components, the granulation (gr) and the supergranulation (sg) and the photon noise. We have zoomed in the top right hand side of the plot the region correspondings to the HIPS.  }
\end{figure}

From the time series of our integrated-Sun-light instruments (with a perfectly regular sampling rate), helioseismologists usually compute the Fourier transform to obtain the power spectrum density (PSD) of the velocity or intensity measurements. This PSD is very rich of information. In Figure~\ref{fig1}, we show the resultant PSD computed with 4768 days of GOLF time series (calibrated into velocity, e.g. \citep{UlrGar2000,GarSTC2005}) by averaging 2382 subseries of 4 days shifted by 2 days (50$\%$ overlapping) and with a duty cycle better than 50$\%$.

In this typical solar PSD we can observe 4 different regions: i) for the lowest frequencies up to $\sim$1 mHz, the spectrum is dominated by the convective noise and it can be modeled by a Harvey-type formula \citep{harvey85} with two components: the granulation and the supergranulation. Moreover, somewhere below $\sim$ 0.45 mHz the g-mode spectrum is buried into this convective noise, ii) between $\sim$1 mHz and the cut-off frequency (defined around $\sim$5.3 mHz e.g. \citep{2005ApJ...623.1215J})  the p modes are clearly visible, iii) above the cut-off frequency down to $\sim$7.5 mHz the region shows an interference pattern of traveling waves usually called: ``High frequency Peaks (HIPS)''  (e.g. \citep{GarPal1998,2003ESASP.517..247C,GarJef1999}), and iv) above $\sim$10 mHz the PSD is dominated by the photon counting noise. 

\begin{figure}[!htb]
 \includegraphics[height=.25\textheight, width=0.45\textwidth]{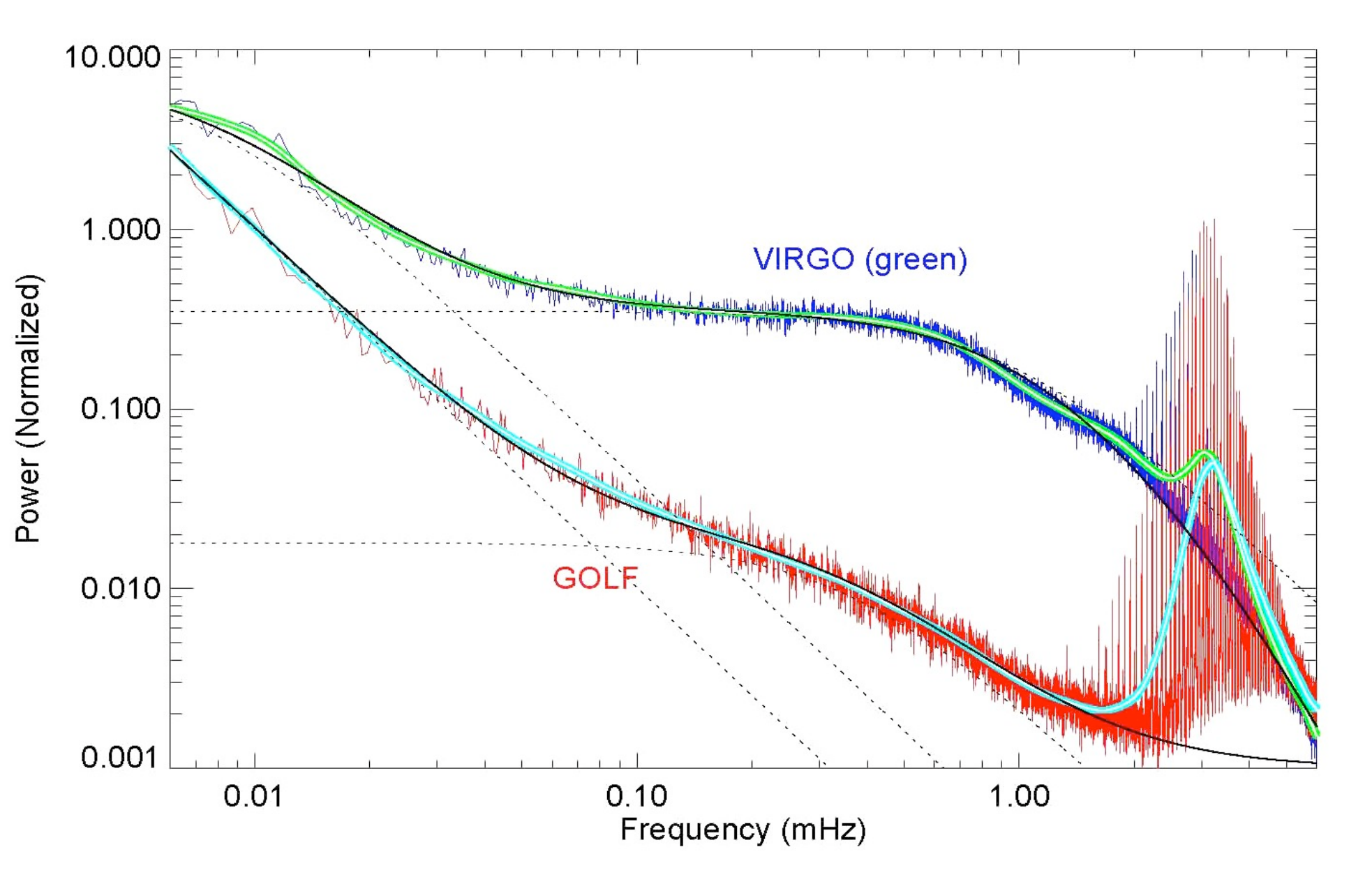}
  \caption{\label{fig2} Comparison between the PSD extracted from velocity (GOLF) or intensity measurements (Virgo/SPM green channel). The fitted background is already showed. (Credits, T. Bedding \& H. Kjeldsen).}
\end{figure}

A comparison between the PSD determined from velocity and intensity measurements from GOLF and VIRGO/SPM is showed in Figure~\ref{fig2} (see e.g. \citep{TouApp1997} for a deeper comparison). To superpose both quantities, we have normalized the PSD to have the same power in the p-mode band. The signal-to-background ratio of the p modes for intensity measurements is not better than 30 while in velocity we reach values above 300.

\section{Extracting p-mode characteristics}

\begin{figure}[!htb]
 \includegraphics[height=.25\textheight, width=0.45\textwidth]{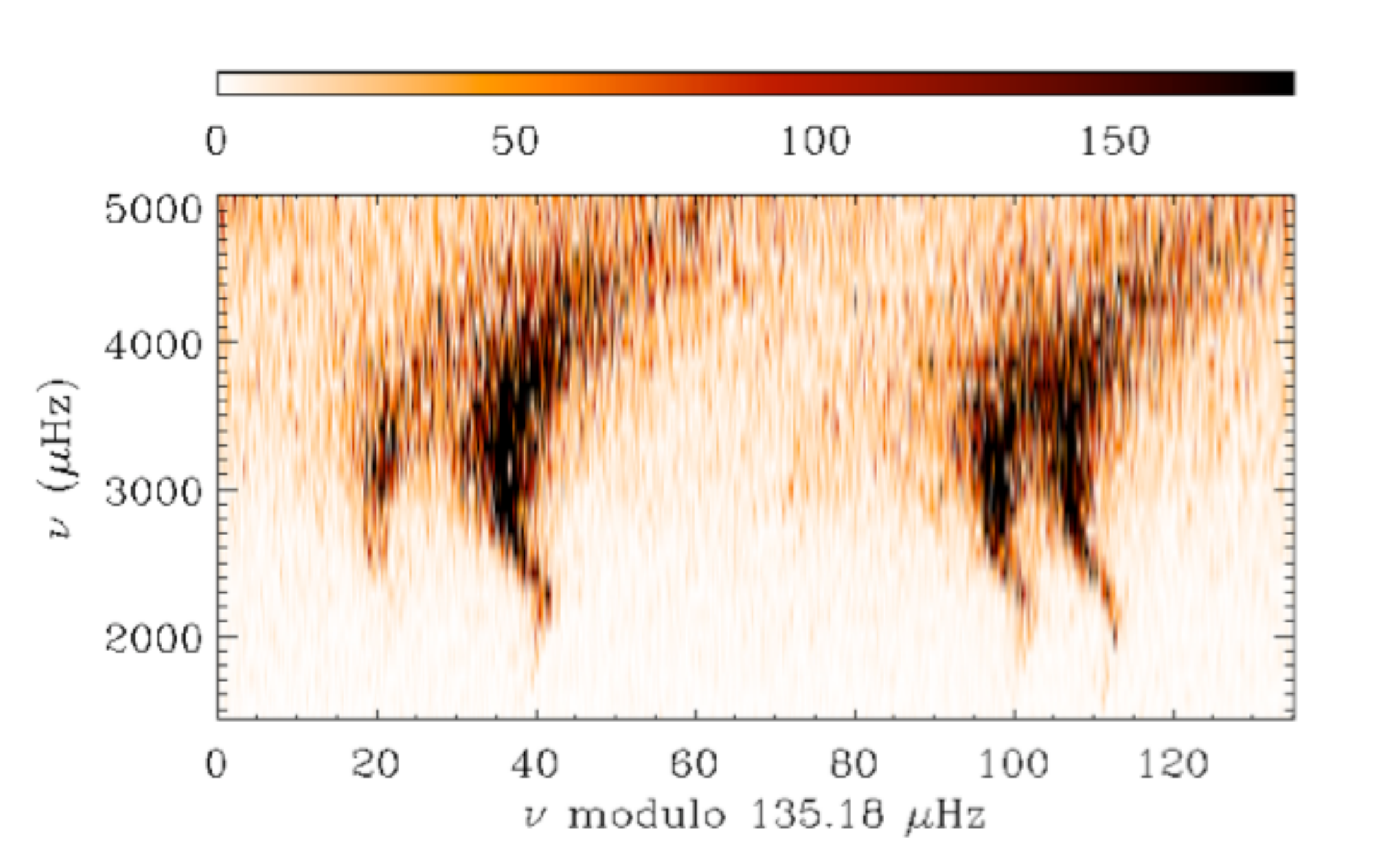}
  \caption{\label{fig3} Echelle diagram built using GOLF velocity measurements with a folding frequency of 135.18 $\mu$Hz.}
\end{figure}

In the solar Sun-as-a-star PSD, we can easily distinguish the low-degree modes up to l=3. Even more, in the case of velocity observations (GOLF or BiSON) we can also detect some l=4 modes in the center of the p-mode range. In Figure~\ref{fig3} we show the echelle diagram \citep{1983SoPh...82...55G} obtained from velocity measurements of GOLF using a folding frequency of 135.18 $\mu$Hz. The ridges of the l=3 \& 1 and l=0 \& 2 are clearly visible in the left- and righ- hand side of the plot respectively. In the middle, between these two pairs of modes, some excess of power can be guessed corresponding to the l=4 modes. 

Compared to the latest solar-like stars observed by CoRoT (see e.g. \citep{Barba09,2009arXiv0907.0608G}) we have no problem to identify the p modes. This is a consequence of the long life-times of such modes at low frequency and the $90^o$ inclination in which we observe the star. Thus, the modes appear very well defined in the PSD.

To obtain the characteristics of the acoustic modes we perform a maximum-likelihood fitting (e.g. \citep{AppGiz1998}). In general this fit is performed by pairs of modes (odd or even degree) separately but very recently, and following what we are used to do in asteroseismology \citep{2008A&A...488..705A}, we have started to do global fittings of the full solar spectrum, including all visible modes and the convective background \citep{1999ESASP.448..135R,2009ApJ...694..144F}. This global technique has been privileged -- in the case of asteroseismology -- because of the high level of correlation between the rotation-inclination angle and the rotational splitting (e.g. \citep{2004ESASP.559..309B,2008A&A...486..867B}), which makes the convergence of the fit more difficult when only a pair of modes is fitted.

Today, the observational limit of low-degree low-order  modes is placed around 1 mHz, with some more radial modes that seem to rise up from the level of noise using GOLF (e.g. \citep{GarReg2001}) or BiSON \citep{2008AN....329..461B} data. 

\section{About p-mode excitation}
It is very well known that global p modes are stochastically excited (e.g. \citep{1977ApJ...212..243G}) yielding to strong and independent temporal variations of the power of the acoustic modes \citep{1998SoPh..181..251C}. However, several groups have reported to have found some correlation between the modes \citep{FogGar1998,1996A&A...311.1024B} while later works could not confirm such correlation  (see e.g. \citep{2004ESASP.559...51B}). However, in this work, a signature of very high energetic events -- such as the so called Sun quake \citep{1998Natur.393..317K}  -- have been unveiled in the summed power of low-degree p modes. 

Recently, it has been reported that solar flares can drive global oscillations in the Sun \citep{2008ApJ...678L..73K} when simultaneous data from the VIRGO package and the X-ray GOES satellite were analyzed (see Figure 2 of \citep{2008ApJ...678L..73K}). Indeed the maximum correlation was reached in the HIPS region. These waves above the cut-off frequency have already been measured in other solar-like stars \citep{2007MNRAS.381.1001K}. If we manage to measure their variation with time during a long period of time in many other solar-like stars (for example through the observations provided by the Kepler satellite \citep{2009IAUS..253..289B}) we would be able to suggest the presence of flares or high-energetic magnetic events in their photospheres. Thus, we could improve our understanding of stellar dynamos.

\section{About the solar activity cycle seen by p modes}
Another way to put constraints on activity cycles and thus, on dynamo processes is through the study of the variation of the p-mode characteristics through the solar-activity cycle. Changes with solar activity were first revealed by \citep{1985Natur.318..449W} and well established later by  \citep{1989A&A...224..253P} using low-degree modes covering a full solar cycle, the  21$^{st}$  from 1977 to 1988.

SoHO has been able to analyze the complete solar cycle 23. Nowadays, we know that all the solar p-mode parameters change with activity (e.g. \citep{2003ApJ...595..446J,2004ApJ...604..969J,2004MNRAS.352.1102C}), even the asymmetry of the modes \citep{JimCha2007}. The shifts induced by the activity cycle are very well correlated with solar activity indexes. Even more, thanks to the old records measured by the first BiSON stations, 3 solar cycles have been seismically analyzed and compared \citep{2007ApJ...659.1749C}.

\begin{figure}[!htb]
 \includegraphics[height=.25\textheight, width=0.45\textwidth]{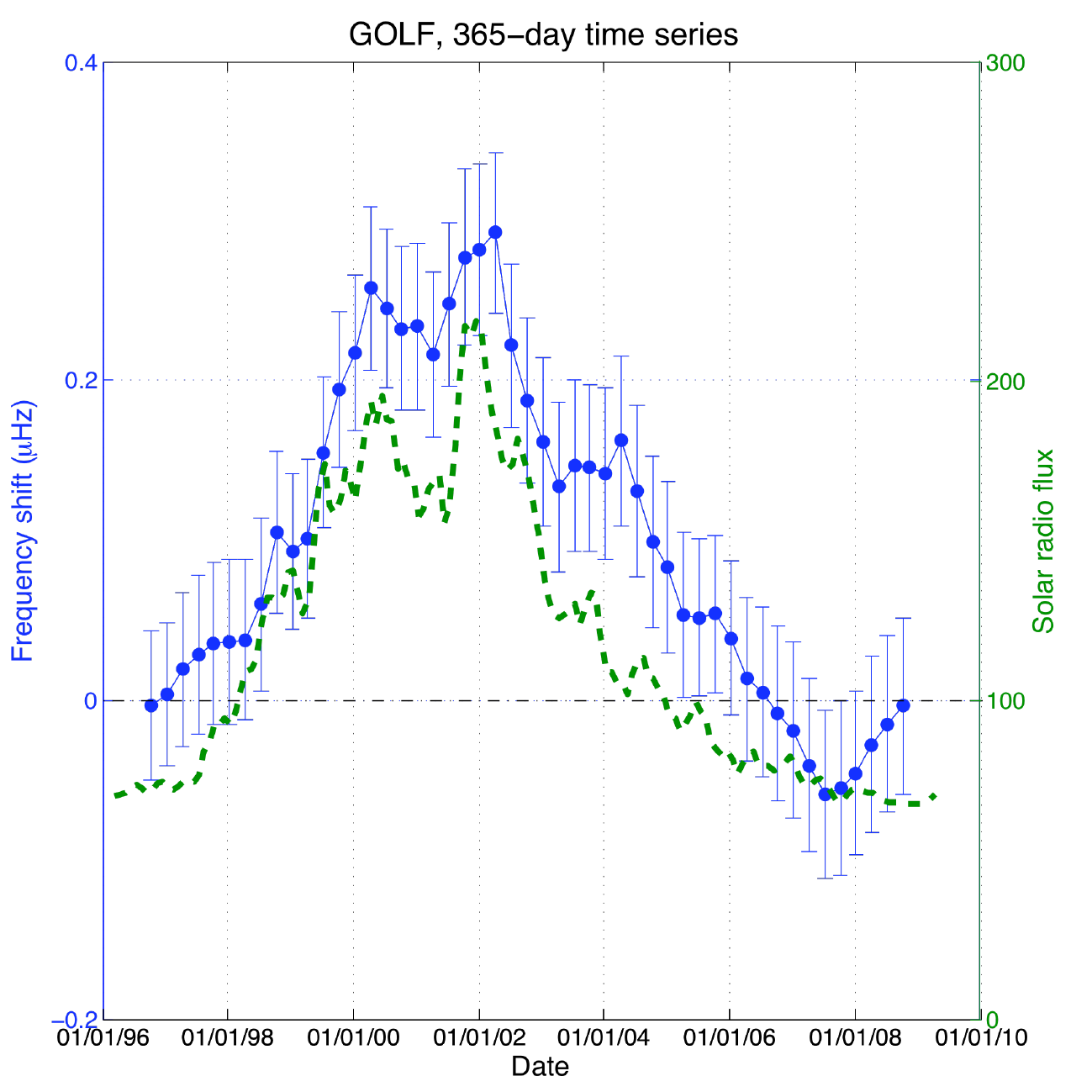}
  \caption{\label{fig4} Average of  the frequency shifts of modes l $\le$ 2 following the methods described in \citep{SalGar09}  using subseries of 365 days (blue points with error bars). We have used the first subseries as the reference one. The green dashed line is the radio flux at 10.7 cm computed using the same subseries.}
\end{figure}

The present solar activity minimum -- between solar cycle 23 and 24 -- is being peculiar. Indeed, all the common activity indicators, such as magnetic indexes (e.g. MPSI\footnote{Magnetic Plage Strength Index \citep{1986ApJ...302L..71C}}) or radiative indexes (e.g. the 10.7-cm radio flux), show a deep and extended solar minima since 2007. However, the analysis of frequency shifts of low-degree modes measured by GOLF \citep{2009arXiv0907.3888S} shows an increase from the second half of 2007, when no significant surface activity was observable (see Figure~\ref{fig4}). Moreover, when the frequency shifts are computed for modes of different degrees separately, the modes l=0 and 2 have a minimum at the second part of the year 2007 and an increase afterwards while the frequency shifts computed using the modes l=1 follow the general decreasing trend of the solar surface activity. This different behavior of the frequency shifts as a function of the mode degree has been interpreted as a consequence of the different geometrical responses to the spatial distribution of the solar magnetic field beneath the solar surface (for more details see \citep{SalGar09}).

\section{Looking for gravity modes}

Gravity modes are the ``master key'' giving access to the core of the Sun. They are very important, for example, for inversions of the rotation profile in the radiative zone. Therefore, coupling them with accurate rotational splittings  obtained form low- and high-frequency low-degree p modes \citep{2008SoPh..251..119G} we could better infer the rotation profile of the solar core \citep{2008A&A...484..517M}.

Gravity modes have been searched since the early days of seismology --back to the 80s-- and several g-mode frequency tables have been published (see for example the review by \citep{HilFro1991}). However none of these candidates have been confirmed with later works. 

Using SoHO and the new ground-based networks, many studies have been published during the last 10 years on this topic (see for example \citep{2008JPhCS.118a2026G,2006soho...18E..22E} and references therein). Starting by looking for peaks above a given threshold (90$\%$ confidence level) the maximum amplitude of g modes was established \citep{AppFro2000,GabBau2002}. Then, to reduce the statistical threshold while maintaining the same statistical significance, multiplets have been searched  \citep{STCGar2004} and some candidates have been found. In particular, the structure detected at 220 $\mu$Hz has been studied in detail, followed in time \citep{2007ApJ...668..594M} and it has been demonstrated that one of its component (the peak at 220.7 $\mu$Hz) --detected in GOLF and VIRGO data-- has been above the level of noise for the last 13 years. Indeed, an extensive analysis of all the housekeeping parameters of the VIRGO package and the pointing of SoHO shows no correlation with this peak implying a solar origin  \citep{JimGar09}. 

Finally, instead of looking for individual peaks, the fingertips of the asymptotic properties has been searched \citep{2007Sci...316.1591G,2008AN....329..476G} using GOLF data. Thus, the signature of the dipole modes has been found with a confidence level in a range 99.5 to  99.99$\%$ not being due to pure noise. Furthermore, a small hint of such a signal was also found in VIRGO/SPM data when the PSD was thresholded \citep{2006ESASP.624E..23G}. The detailed analysis of the detected signal has given some new light on the structure and dynamics of the solar core. On one side, it favors a faster rotation rate in the core (3 to 5 times faster than the rest of the radiative region). On the other hand, it tends to favor \citep{2008SoPh..251..135G} solar standard models computed using old surface abundances in comparison to those computed using the new ones \citep{2005ASPC..336...25A}. It is important to notice that both results were obtained without taking into account p modes.

The detection of the asymptotic signature of the dipole modes and the latest predictions about their surface amplitudes (around a couple of mm/s \citep{2009A&A...494..191B}) could mean that these modes could be detected individually in a very near future.

\section{Conclusions}

A very exciting future is ahead us in helioseismology. From the understanding of the solar core of the Sun to their excitation  mechanism or the physical processes behind the dynamo and the solar activity cycle, the detection of more and more precise low-degree modes will be very important. Moreover, present asteroseismic missions, such as CoRoT, have already shown that models based on the Sun alone, have failed when they are applied to other Solar-like stars \citep{2008Sci...322..558M}. New results based on this satellite as well as on the NASA mission Kepler will provide more constraints for a wider solar-like stars in many different conditions. The sum of all these observations will definitely impose new constraints in the modeling of solar-like stars.


\begin{theacknowledgments}
 The GOLF, SOI/MDI, and VIRGO instruments onboard SoHO are the cooperative effort of many individuals to whom we are indebted. SoHO is a project of international collaboration between ESA and NASA. This work was supported by the CNES-GOLF grant at the Service d'Astrophysique, CEA-Saclay.
 \end{theacknowledgments}

\bibliographystyle{aipproc}   

\bibliography{/Users/rgarcia/Desktop/BIBLIO} 

\IfFileExists{\jobname.bbl}{}
 {\typeout{}
  \typeout{******************************************}
  \typeout{** Please run "bibtex \jobname" to optain}
  \typeout{** the bibliography and then re-run LaTeX}
  \typeout{** twice to fix the references!}
  \typeout{******************************************}
  \typeout{}
 }

\end{document}